\documentclass[a4paper,12pt]{article}
\usepackage[utf8]{inputenc}
\usepackage{cancel}
\usepackage{ulem}
\usepackage{amsfonts}
\usepackage{amssymb}
\usepackage{graphicx}
\usepackage{amsmath}
\usepackage{enumerate}
\usepackage{mathtools}
\usepackage{subfig}
\usepackage{color}
\usepackage{tikz}
\usepackage{float}
\usepackage{here}
\usepackage{cite}
\usepackage{mathrsfs}
\usepackage{float,epsfig}
\usepackage{dcolumn}
\usepackage{lscape}
\usepackage{subfig}
\usepackage{graphicx}
\usepackage{bm}
\usepackage{amsmath,amssymb,amsthm}
\usepackage[colorlinks=true,linkcolor=blue,citecolor=red]{hyperref}
\usepackage{multirow}

\usetikzlibrary{calc}
\newcommand{\be}{\begin{equation}}
\newcommand{\ee}{\end{equation}}
\newcommand{\bea}{\setlength\arraycolsep{2pt} \begin{eqnarray}}
\newcommand{\eea}{\end{eqnarray}}

\def\0{{\sst{(0)}}}
\def\1{{\sst{(1)}}}
\def\2{{\sst{(2)}}}
\def\3{{\sst{(3)}}}
\def\4{{\sst{(4)}}}
\def\5{{\sst{(5)}}}
\def\6{{\sst{(6)}}}
\def\7{{\sst{(7)}}}
\def\8{{\sst{(8)}}}
\def\sst#1{{\scriptscriptstyle #1}}

\makeatletter \@addtoreset{equation}{section}

\begin{document}

\title{{\normalsize \textbf{\Large  Thermodynamics and Criticality   of Noncommutative RN--AdS Black Holes }}}
\author{ \small   Wijdane El Hadri$^{1,2}$ \footnote{w.elhadri@uiz.ac.ma} 	and   Maryem  Jemri$^3$ \footnote{Corresponding author: maryem.jemri@um5r.ac.ma}  \thanks{
Authors are listed  in alphabetical order.} \hspace*{-8pt} \\
%EndAName
{\small $^1$ENSIASD, Taroudant,  Ibnou Zohr University, Agadir, Morocco  }\\
{\small $^2$LPTHE, Faculty of Science, Ibnou Zohr University, Agadir, Morocco }\\
{\small $^3$ESMaR, Faculty of Science, Mohammed V University in Rabat, Rabat, Morocco  } }
\maketitle

\begin{abstract}
Inspired by string theory topics, we  investigate the Reissner--Nordström--AdS  black holes in noncommutative geometry with  Lorentzian-smeared distributions.
 Concretely,  we study certain thermodynamic properties including the criticality  behaviors
by computing the relevant quantities. For large radius approximations,  we   first derive the asymptotic expansions of the mass and charge functions   appearing  in the  metric function of such black holes.  Then, we approach  the thermodynamical behavior in the extended phase space.  After the  stability discussion, we inspect    the $P$--$V$ criticality  in  noncommutative  geometry  by calculating the corresponding   thermodynamic quantities. As a result, we show  that the proposed black holes exhibit certain   similarities with   Van
der Waals fluid systems.  Finally, we   present a discussion  on  the Joule–Thomson expansion showing   perfect universality  results appearing in  charged AdS black holes.

{\noindent} 

\end{abstract}
\newpage
\tableofcontents
\newpage
\section{Introduction}
The physical properties of our universe raise strange and mysterious concepts, among which is that of black holes \cite{1}. They are considered to be the most fascinating and captivating objects\cite{2}. Black holes were predicted by general relativity and are theoretically modeled as solutions to Einstein  equations \cite{3}. They have recently become observable objects of great interest. To understand the different behaviors of black holes, numerous connections with various theories and several fields of research have been considered\cite{4,up8,up5,up6}. With regard to black holes,  the Hawking  area theorem, which asserts that the surface area of a black hole's event horizon can never decrease, has opened up a new avenue of research for understanding black holes from a thermodynamic point of view. In this context, the surface area of the horizon has been associated with entropy,  and the surface gravity has been interpreted as temperature. Moreover, an analogy has been established between the four laws of thermodynamics and the mechanics of black holes \cite{up12,5,6,7,8,9,10,up15}. This enables the consideration of both the thermodynamic and quantum aspects of black holes \cite{up9,11,up11,up13,up14,up16}.

Recently, the thermodynamics of many classes of black holes in Anti-de Sitter (AdS) spacetime has also been explored in \cite{12}. In this spacetime, the negative cosmological constant has been treated as a thermodynamic variable and considered as a pressure in the equation of state \cite{13}. This cosmological constant induces phase transition phenomena that appear naturally in black hole physics \cite{12}.  The Hawking-Page phase transition which  has been defined as a first-order phase transition between thermal radiation and large black holes has been investigated \cite{14}. In spacetime with arbitrary dimensions, a comparison  has been made between the phase transition behavior of charged AdS black holes and that of Van der Waals fluids \cite{77}. 

A new development in noncommutative spacetime field theory presents a simplified methodology for reproducing string  theory phenomenology, particularly in the low-energy approximation. Noncommutative geometry (NCG) provides a framework for quantized spacetime, expressed by commutation relations that go beyond those appearing in quantum mechanics \cite{r6,r7,r8,r9,r10,r11}.  Concretely, this geometry has been extensively studied in connection with type II superstrings   in the presence of D-branes where the  noncommutativity parameters  have been linked to  the the antisymmetric B field \cite{15}. In the context of charged black holes,  the noncommutativity allows both mass and charge to be modified, which compensates for the classical singularities of the Reissner-Nordstrom solution \cite{up3,up4}. Previous studies have revealed significant developments concerning noncommutative effects in gravity \cite{16,166,30,i5}. In this approach, the source term for matter is modified while the Einstein tensor in the field equations remains unchanged. More specifically, the usual point mass in  the Einstein equations is replaced by a spread Gaussian or Lorentzian distribution.  In this context, many aspects of black hole physics have been studied, including Hawking temperature and tunneling processes \cite{160,up0,up1,up2}, shadow properties \cite{18,19,20,up10,17}, topological features in modified gravity theories \cite{21}, gravitational lensing \cite{22}, and matter accretion \cite{23}. In addition, a new approach to integrating non-commutativity into gravitational scenarios has been suggested, treating it as a perturbative effect \cite{24}.

The objective of this work is to take part in these activities by exploring Reissner-Nordström-AdS black holes in a noncommutative spacetime with spread Lorentzian distributions.
 Concretely,  we  examine certain thermodynamic properties including the critical behaviors
by computing the relevant quantities. We start by   deriving  the asymptotic expansions of the mass and charge functions for large radii  appearing  in the  metric function of such black holes needed to  approach  the thermodynamical aspect in the extended phase space.  After the  stability discussion, we  investigate    the $P$--$V$ criticality  in  noncommutative  geometry  by calculating the  associated thermodynamic quantities. As a result, we show  that the proposed black holes exhibit   certain similarities with   Van
der Waals fuild systems.  Finally, we  provide  a discussion  on  the Joule–Thomson expansion  revealing   perfect universality  findings  appearing in  charged AdS black holes in ordinary spacetimes.

The structure of this paper is as follows.  In section 2, we  build  the Reissner--Nordström--AdS  black holes in noncommutative spacetime with  Lorentzian-smeared distributions.  In section 3, we compute and analyze certain  thermodynamics quantities needed to approach stability behaviors.  In section 4, we  investigate    the criticality  and make contact with Van der Waals fluid  systems.   In section 5, we examine   the Joule-Thomson expansion effects.   In the last section, we  present    concluding   remarks.

\section{Noncommutative RN--AdS black holes}
Motivated by applications of noncommutative geometry to certain physical  models   in the context of  string theory, we would like to   propose   a NC--RN--AdS black hole metric.  It  is recalled that  noncommutative geometry  provides a framework for describing quantum gravitational effects at small scales. In this way,    the spacetime coordinates become operators obeying the following relations
\begin{equation}
[x^\mu , x^\nu] = i\,\theta^{\mu\nu}. 
\end{equation}
In these relations, the quantity  $\theta^{\mu\nu}$  represents   a constant antisymmetric tensor where  one has used $\hbar=1$.  This geometry has been extensively investigated in connection with several string theory topics including D-brane physics where such a tensor has been related to the $B_{\mu\nu}$   stringy field   \cite{15}.    Roughly  speaking,  any  NC parameter $\Theta$ introduces a minimal length scale $\sqrt{\Theta}$, smoothing out classical singularities.
Applications of NCG to black holes reveal modifications in horizon structures, thermodynamics, quasinormal modes, and shadows.  Concretely,  RN--AdS black holes in noncommutative geometry  will display novel thermodynamic behaviors in the extended phase space, where the cosmological constant  will be  interpreted as a pressure term.

%This work focuses on the thermodynamics and $P$--$V$ criticality of RN--AdS black holes with Lorentzian-smeared mass and charge distributions, highlighting the effects of the NC parameter $\Theta$.
In the present work, we assume that the metric of a static, spherically symmetric RN--AdS black hole in noncommutative geometry reads as 
\begin{equation}
 ds^2 = -f(r) dt^2 + \frac{dr^2}{f(r)} + r^2 d\theta^2+  r^2 \sin^2 \theta d\phi^2.
\end{equation}
In this way,  the deformed  metric function can have the following general form 
\begin{equation}
 f(r) = 1 - \frac{2m(r)}{r} + \frac{q^2(r)}{r^2} - \frac{\Lambda r^2}{3},
\end{equation}
where  $ \Lambda$ is the cosmological constant.  The radial functions $m(r)$ and $q(r)$ will be specified later on. 
%\subsection{Einstein Equations and Lorentzian Distribution}
Indeed, the explicit expression of the  metric function $f(r)$ can be obtained by solving the Einstein equations involving a cosmological constant $\Lambda$
\begin{equation}
 G_{\mu\nu} + \Lambda g_{\mu\nu} = 8\pi T_{\mu\nu},
\end{equation}
where  $ G_{\mu\nu}$ is the Einstein  tensor.  The energy-momentum tensor $T_{\mu\nu}$  is constructed from a Lorentzian mass and charge  densities 
\begin{equation}
 \rho_M(r)= \frac{M \sqrt{\Theta}}{\pi^{\frac{3}{2}} (r^2 + \Theta)^2},  \qquad   \rho_Q(r)= \frac{Q \sqrt{\Theta}}{\pi^{\frac{3}{2}} (r^2 + \Theta)^2}
\end{equation}
where $\Theta$ is the noncommutative parameter of dimension  $[L^{2}]$ \cite{26,r1,r2,r3,r4,r5}. $Q$ indicates the  black hole charge,  and  $M$ is the total mass diffused throughout the region of linear sizes $\sqrt{\Theta}$ \cite{ll,ll1}. Thus, the smeared mass  and charge distribution functions take the form
\begin{equation}
\begin{aligned}
m(r) &= \int_{0}^{r} \rho_{M}(r) 4\pi r^{2}  dr\\ q(r) &= \int_{0}^{r} \rho_{Q}(r) 4\pi r^{2}  dr.
\end{aligned}\label{0ll}
\end{equation}
This smooth distribution replaces the point-like source in classical RN--AdS solutions leading to regularized metric functions $m(r)$ and $q(r)$ which should  be   computed using certain approximations.   The smeared mass and  the charge functions are given by a Lorentzian profile as follows
\begin{equation}
\begin{aligned}
m(r) &= \frac{2M}{\pi} \arctan\!\left(\frac{r}{\sqrt{\pi \Theta}}\right) 
       - \frac{2M \sqrt{\Theta}}{\sqrt{\pi}} \frac{r}{r^2 + \pi \Theta}, \\
q(r) &= \frac{2Q}{\pi} \arctan\!\left(\frac{r}{\sqrt{\pi \Theta}}\right) 
       - \frac{2Q \sqrt{\Theta}}{\sqrt{\pi}} \frac{r}{r^2 + \pi \Theta}.
\end{aligned}\label{ll}
\end{equation}
To handle such  expressions,  calculation techniques   and certain approximations will be considered.  Indeed, we 
take  $\alpha=\pi\Theta$ and $y=r/\sqrt{\alpha} \gg 1$.  Using   the following  expansions
\begin{equation}
 \arctan y = \frac{\pi}{2} - \frac{1}{y} + \frac{1}{3y^3} + O\left(\frac{1}{y^5}\right),\quad
 \frac{r}{r^2+\alpha} = \frac{1}{r} - \frac{\alpha}{r^3} + O\left(\frac{\alpha^2}{r^5}\right). \label{kk}
\end{equation}

Inserting these  expansions into Eq.(\ref{ll}), we obtain
\begin{equation}
m(r) = \frac{2M}{\pi} \left( \frac{\pi}{2} - \frac{\sqrt{\pi}\Theta}{r} + \frac{(\pi\Theta)^{3/2}}{3r^3} \right) - \frac{2M\sqrt{\Theta}}{\sqrt{\pi}} \left( \frac{1}{r} - \frac{\pi\Theta}{r^3} \right) + O\left( \frac{\Theta^{5/2}}{r^5} \right).
\end{equation}
The calculation leads to
\begin{equation}
m(r) = M - \frac{4M\sqrt{\Theta}}{\sqrt{\pi}r} + \frac{8M\sqrt{\pi}}{3} \frac{\Theta^{3/2}}{r^3} + O\!\left( \frac{\Theta^{5/2}}{r^5} \right).
\end{equation}
Similarly, for the electric charge, we get 
\begin{equation}
q(r) = Q - \frac{4Q\sqrt{\Theta}}{\sqrt{\pi}r} + \frac{8Q\sqrt{\pi}}{3} \frac{\Theta^{3/2}}{r^3} + O\!\left( \frac{\Theta^{5/2}}{r^5} \right).
\end{equation}

Substituting these expansions into $f(r)$ yields
\begin{equation}
 f(r) = 1 - \frac{2M}{r} + \frac{8M\sqrt{\Theta}}{\sqrt{\pi} r^2} + \frac{Q^2}{r^2}  - \frac{8Q^2 \sqrt{\Theta}}{\sqrt{\pi} r^3}  - \frac{\Lambda r^2}{3}+ O(\Theta^{3/2}).
\end{equation}
To simplify the computations, we use a new parameter 
  $a$ with dimension of $[L]$
\begin{equation}
 a= \frac{8\sqrt{\Theta}}{\sqrt{\pi}}
\end{equation}
carrying the NC modification in the thermodynamic quantities.
 In this context, the   RN--AdS black hole  metric function in noncommutative geometry   involves the following form
\begin{equation}
 f(r) = 1 - \frac{2M}{r} + \frac{aM}{ r^2} + \frac{Q^2}{r^2} - \frac{aQ^2} {r^3}- \frac{\Lambda r^2}{3}.
\end{equation}
Taking   $Q=0$, we obtain
\begin{equation}
 f(r) = 1 - \frac{2M}{r}  + \frac{aM}{ r^2}   - \frac{\Lambda r^2}{3},
\end{equation}
representing the  metric function  of a Schwarzschild-AdS  black hole  in noncommutative spacetime 
geometry \cite{30}.

For a comprehensive thermodynamic analysis, we  consider  a nonzero cosmological constant $\Lambda$, allowing one to examine the critical behavior and Joule–Thomson expansion in the noncommutative Schwarzschild–AdS backgrounds.

\begin{figure}[h!]
    \centering
    \begin{tabular}{cc}
       \includegraphics[width=7.5cm,height=7.5cm]{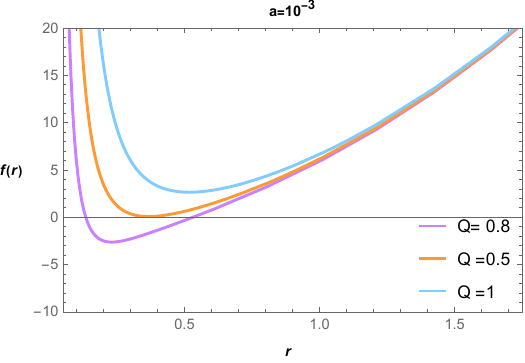} & 
         \includegraphics[width=7.5cm,height=7.5cm]{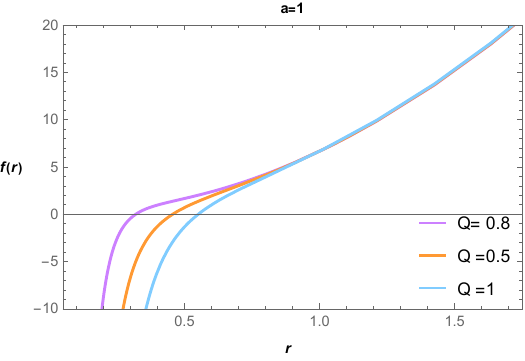} 
    \end{tabular}
     \caption{Effect of the charge parameter $Q$ and  $a$ on the metric function $f(r)$ for $M=1$ and $\Lambda=-20$.}
    \label{Fig3.1}
\end{figure}

Before studying the thermodynamical properties of these solutions, we first  examine  the behavior of the black hole metric function. By fixing the mass and the cosmological constant, the analysis is carried out in terms of the two main parameters $(a,Q)$. Fig.(\ref{Fig3.1}) roughly shows  such behaviors.

For a fixed value of 
$a$,   it has been observed that there exists a critical charge 
$Q_c$ corresponding to a double root of 
$f(r)=0$, which gives an extremal black hole solution. In fact, the spacetime can present two horizons (inner and outer) or a naked singularity. The latter  appears  for $Q>Q_C$. However,    the solution describing  a non-extremal black hole is assured when $Q<Q_C$.

\section{Thermal and stability aspects of RN--AdS black holes  in noncommutative geometry   }
  To inspect  some  physical  behaviors including the stability,  one should compute   relevant thermodynamic quantities. In this section,  we calculate    the mass, the temperature and the heat capacity.  In particular, we explore the  thermal and  the stability behaviors in the presence of such a spacetime modification using noncommutative geometry  techniques.
  
  \subsection{Thermal behaviors }
\begin{figure}[h!]
    \centering
       \includegraphics[width=12cm,height=8cm]{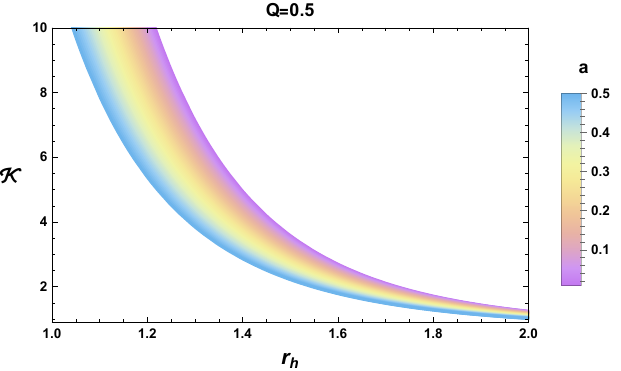} 
    \caption{Profile of the embedding diagram within different values of the correction parameter $a$.}\label{k}
    \label{Fig3.2}
\end{figure}

We start by computing the  mass of   the noncommutative RN--AdS black holes by solving the constraint   $f(r_h)=0$  where  $r_h$ denotes the horizon radius.  It has been observed that an explicit expression for such a radius is not a simple task  when $\Lambda \neq 0$. However, taking $\Lambda = 0$, we can obtain a solution for $f(r_h)=0$ by considering  small valus for $Q$ and $\Theta$. Indeed, one gets
\begin{equation}
r_h = 2M - \frac{Q^2}{2M} - \frac{a}{2} - \frac{a^2}{ 8M} - \frac{a^3}{16 M^2} - \frac{128a^4}{5 M^3}.
\end{equation}
At this point, we would like to provide three comments.
First, puting now  $a=0$, we obtain
\begin{equation}
r_h = 2M - \frac{Q^2}{2M}, 
\end{equation}
  recovering the RN horizon radius by taking small valus limit of the charge in the expresion \begin{equation}
r_h = M+\sqrt{M^2-Q^2}. 
\end{equation} 
Second, ignoring higher powers, the positivity of 
$r_h$ requires $4M^2-aM>Q^2$. Third, It would be interesting to examine how the NC parameter influences the appearance of singularities or contributes to the regularization of the black hole at $r=0$. To address this point, a  geometric description  is needed. It recalled that the Kretschmann scalar, a curvature invariant, is used to characterize the intensity of the curvature of spacetime at a given point. Indeed, it  is specified as a special contraction of the components of the Kretschmann tensor
\begin{equation}
\mathcal{K} = R_{abcd} R^{abcd},
\end{equation}
The calculation leads to
\begin{equation}
\mathcal{K} = \frac{8} {3 r^{10}} \left( X
+Y 
+ Z\right),
\end{equation}
where one has used
\begin{equation}
\begin{aligned}
X=&69 a^{2} Q^{4} - 75 a Q^{2} (a M + Q^{2}) r \\
Y=&  3 (7 a^{2} M^{2} + 34 a M Q^{2} + 7 Q^{4}) r^{2}\\
Z=& 18 M^{2} r^{4} 
+ a \Lambda Q^{2} r^{5} + \Lambda^{2} r^{10}-36 M (a M + Q^{2}) r^{3}  
\end{aligned}
\end{equation}
In Fig.\ref{k}, we depict the behavior of such a quantity in terms of the radial coordinate $r$ for
various ranges  of the NC parameter $a$.
It has been remarked, from the figure,  that
the Kretschmann number increases as $a$ decreases. This indicates that spacetime is significantly
curved, leading to deviations from the flat spacetime for the high values of the NC parameter.  The mass is  found to be 
\begin{equation}
M =\frac{\Lambda  \,r_h^{5}+3 Q^{2} a -3 Q^{2} r_h -3 r_h^{3}}{3 r_h \left(a-2 r_h  \right)}.    
\end{equation}
Removing the electric charge  $Q=0$, we recover the expression found in \cite{30} being 
\begin{equation}
M =\frac{\Lambda  \,r_h^{4}-3 r_h^{2}}{ 3(a-2 r_h) }.    
\end{equation}
To  obtain the  Hawking temperature\cite{i1}, we exploit the relation 
\begin{equation}
T_H=  (4\pi)^{-1}\left. \frac{ \mathrm{d}f(r) }{ \mathrm{d}r } \right|_{r = r_h} .    
\end{equation}
The computations give 
\begin{equation}
T_H = \frac{-4 \Lambda  a \,r_h^{5}+6 \Lambda  \,r_h^{6}+3 Q^{2} a^{2}-12 Q^{2} a r_h +6 Q^{2} r_h^{2}+6 a \,r_h^{3}-6 r_h^{4}}{12 \pi \left(a-2 r_h  \right) r_h^{4} }.  \label{xxx} 
\end{equation}
 Taking $Q$ = 0, the  Hawking temperature can by reduced to
\begin{equation}
T_H=\frac{3 \Lambda  \,r_h^{6}-3 r_h^{4}+3 a \,r_h^{3}-2 \Lambda  a \,r_h^{5}}{6 \left(a-2 r_h  \right) r_h^{4} \pi},
\end{equation}
recovering the result obtained in \cite{30}.  Considering  $a= 0$ and $\Lambda= 0 $,  we  recover  the  temperature of the ordinary Schwarzschild black
hole given by  $T_{H}=\frac{1}{4\pi r_{h}}$ \cite{TT}.
In order to illustrate the behavior of the temperature, we may plot the
 above expression in terms of the event horizon radius, by restricting to
 suitable regions of the reduced moduli space by considering the cosmological constant. Fig.(\ref{Fig3.2}) depicts such a 
 thermal behavior.

 \begin{figure}[h!]
    \centering
    \begin{tabular}{cc}
       \includegraphics[width=8cm,height=7cm]{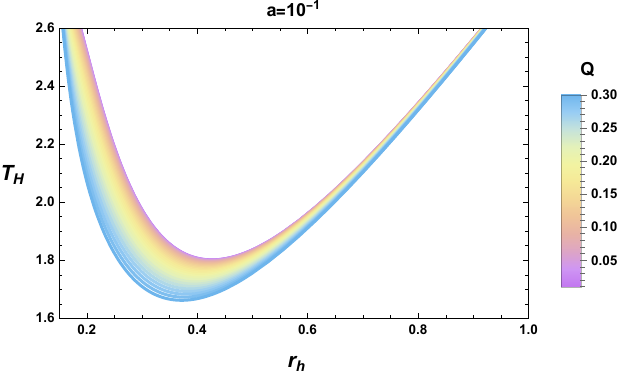} & 
         \includegraphics[width=8cm,height=7cm]{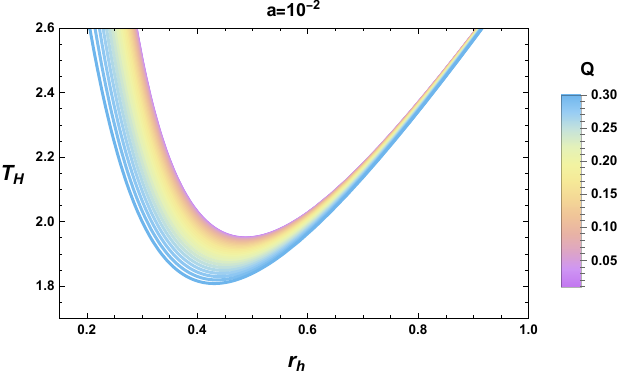} 
    \end{tabular}
    \caption{Effect of the charge parameter $Q$ and $a$ on the Hawking temperature $T$  as a function of $r_h$ by taking  $\Lambda=-50$.}
    \label{Fig3.2}
\end{figure}

This  figure  reveals  that the Hawking temperature  decreases  to a minimal  value. Then,  it increases.  An examination shows that 
the minimal  value decreases   by augmenting  the electric charge $Q$. It has been observed that  the increasing of the parameter  $a$ leads to  small values of the Hawking temperature.

\subsection{Stability behaviors}
\begin{figure}[h!]
    \centering
    \begin{tabular}{cc}
       \includegraphics[width=7.5cm,height=7.5cm]{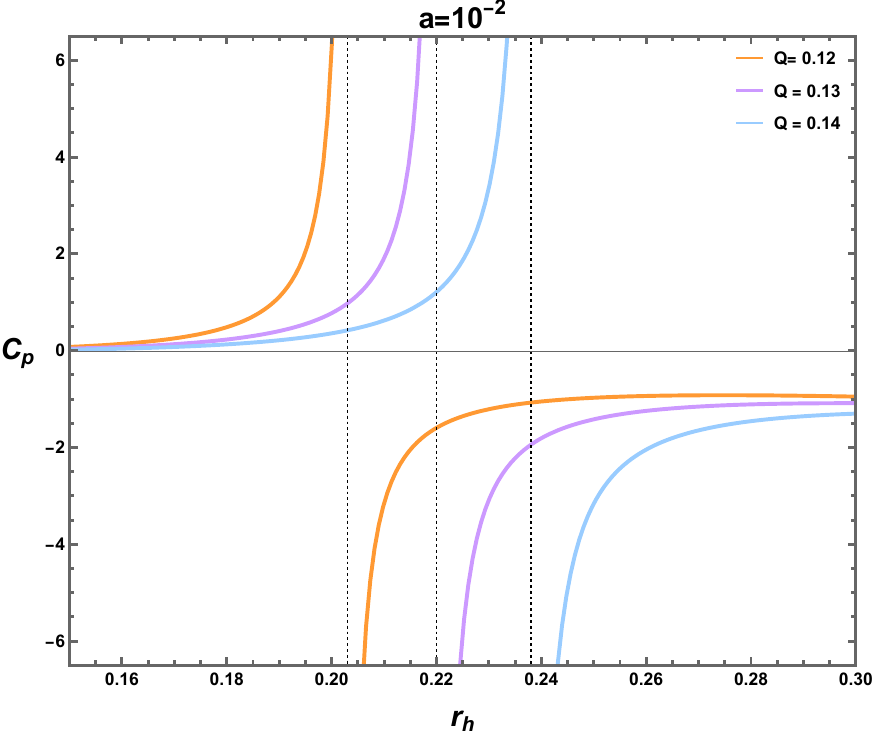} & 
         \includegraphics[width=7.5cm,height=7.5cm]{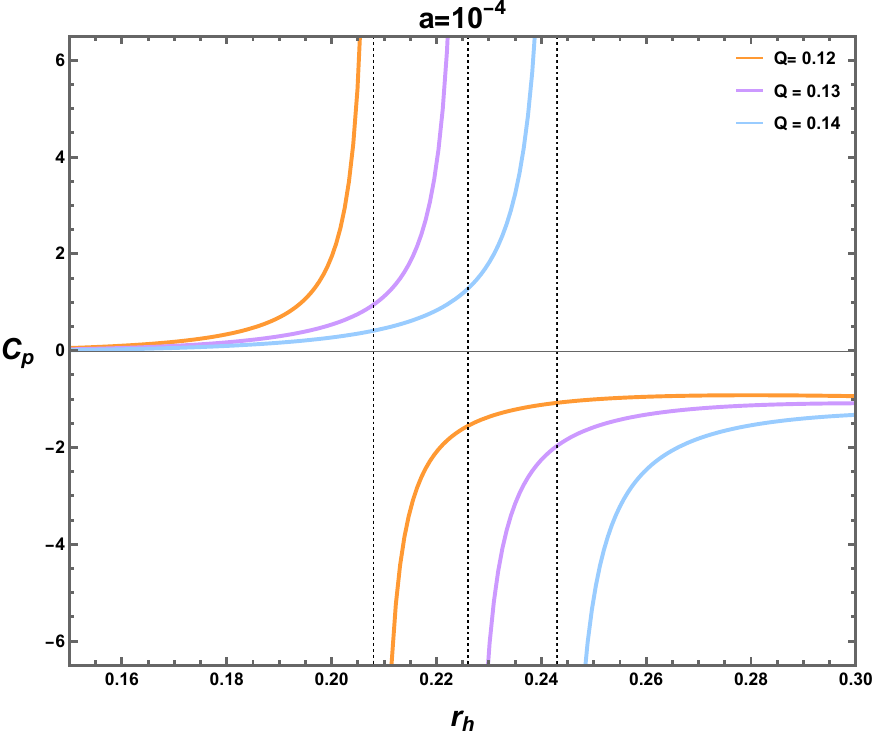} 
    \end{tabular}
    \caption{Effect of the charge parameter $Q$ and  the NC  parameter $a$ on the heat capacity as a function of  $r_h$  for $\Lambda=-0.1$.}
    \label{4}
\end{figure}

The local thermodynamic stability of  black holes  can be approached  via the heat capacity  relation \begin{equation}
C_p = T_H  \frac{\partial S}{\partial T_H}.
\end{equation}
To compute such a  quantity, we  need to determine first  the entropy via  the Bekenstein--Hawking area law
\begin{equation}
S =\dfrac{\cal
A}{4} 
\end{equation}
 where ${\cal A}=\iint\sqrt{g_{\theta \theta}g_{\phi \phi}} \, d\theta \, d\phi=4\pi r^2_h $
is the surface area
of the black hole event horizon. By help of   the Hawking temperature $T_H$ given in Eq.(\ref{xxx}), we find that the heat capacity can be given by 
{\footnotesize
\begin{equation}
C_p =\frac{6 \pi  \left(-\frac{4}{3} \Lambda  a \,r_h^{5}+2 \Lambda  \,r_h^{6}+Q^{2} a^{2}-4 Q^{2} a r_h +2 Q^{2} r_h^{2}+2 a \,r_h^{3}-2 r_h^{4}\right) r_h^{3}}{6 \Lambda  \,r_h^{7}-6 \Lambda  a \,r_h^{6}+2 \left(\Lambda  \,a^{2}+3\right) r_h^{5}-12 a \,r_h^{4}+3 \left(-6 Q^{2}+a^{2}\right) r_h^{3}+54 Q^{2} a \,r_h^{2}-33 Q^{2} a^{2} r_h +6 Q^{2} a^{3}}.
\end{equation}}
Taking $a=0$ and $Q=0$, we get 
\begin{equation}
C_p =\frac{2 \pi  \,r_h^{2} \left(\Lambda  \,r_h^{2}-1\right)}{\Lambda  \,r_h^{2}+1}
\end{equation}
recovering the standard AdS--Schwarzschild black hole expression\cite{266}. Based on the sign of the heat capacity, we can identify the stability of the assosiated black hole solutions. Indeed, a locally stable thermodynamic system can occur if $C_p>0$, while an unstable solution arises if $C_p<0$. A graphical depiction of this phenomenon is illustrated in Fig.(\ref{4}), in which we plot $C_p$ as a function of $r_{h}$ for certain  points in the moduli space.

For a given point in the parameter space, we observe that the heat capacity curves
are disconnected at the critical values
$r_{h}=r_{h}^{c}$ associated with the minimum temperature. Fixing the charge,  it has been observed that  $r_{h}^{c}$  increases by decreasing the NC parameter $a $. Moreover, 
it has been  remarked that  two separated branches appear showing  the proposed models  develop  a black hole transition from
a stable phase to an unstable one, specified  by $r_h<r_{h}^{c}$ and $r_h>r_{h}^{c}$, respectively. This  divergence  supports  a second
order phase transition which will be discussion in the next section.

\section{$P$-$V$ criticality  and  phase transitions }
In this section, we    would like to study    the  $P$-$V$ criticality  and  phase transitions  of noncommutative RN--AdS black holes. To do so, certain    thermodynamic  quantities should be computed.
 \subsection{$P$-$V$ criticality} 
\begin{figure}[h!]
    \centering
    \begin{tabular}{cc}
       \includegraphics[width=7cm,height=7cm]{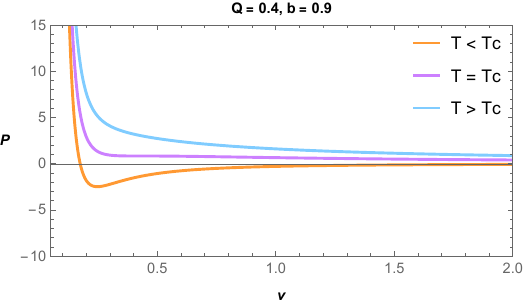} & 
         \includegraphics[width=7cm,height=7cm]{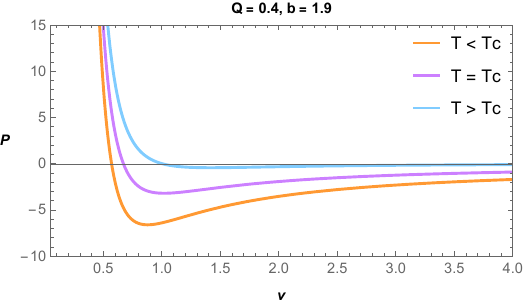} 
    \end{tabular}
\caption{Pressure in terms of $v$ for different values of $T$ and $b$ with $Q=0.4$. }
\label{F41}
\end{figure}

To start, we need to  establish the  thermodynamic equation of state. In the extended phase space,  the cosmological constant $\Lambda$ is  considered as a thermodynamic variable \begin{equation}
P = -\frac{\Lambda}{8\pi}.
\end{equation}
This approach not only provides a more complete thermodynamic description, but also promotes the emergence of rich phase structures and critical phenomena similar to those found in ordinary thermodynamic systems, such as Van der Waals fluids. This elaboration is the starting point for determining the equation of state used to verify the $P$-$V$ criticality of the system \cite{31,32,33,34,340,35}.  Computations lead to 
\begin{equation}
 P=\frac{12 \pi  T r_h^4 \! \left(2 r_h -a \right)+6 r_h^3 \! \left(a -r_h \right)+3 Q^2 \! \left(a^{2}-4 a r_h +2 r_h^{2}\right)}{16 r_h^{5} \left(3 r_h -2 a \right) \pi}\label{P}
\end{equation}
Removing the electric charge  $Q=0$, we recover the expression found in \cite{30} being 
\begin{equation}
 P =\frac{3\left(2 \pi T a r_h - 4 \pi T r_h^{2} - a + r_h\right)}{8 r_h^{2} (2a - 3r_h)\pi}.
\end{equation}
To obtain the thermodynamic critical  values, we need to determine the black hole thermodynamic volume. Indeed, it is given by 
 \begin{equation}
 V=\frac{4 \pi  \,r_h^{3}}{3}.\label{V}
 \end{equation}
A first sight,  the computations  of the critical quantities look like a hard task.  However,  we can use the techniques   explored in   \cite{37,38}.  Considering  $\dfrac{2a}{3r_h}$ as a constant   $b$   as follows \begin{equation} 
\dfrac{2a}{3r_h}=b,
\end{equation}
such critical values could be approached where certain conditions should be imposed to  get acceptable quantities.   Solving the constraints
\begin{equation}
\frac{\partial P}{\partial v }=0,\hspace{1.5cm}\frac{\partial ^{2}P}{%
\partial v^{2}}=0,
\end{equation}
the critical thermodynamic  quantities are shown to be 
\begin{eqnarray}
P_c &=& \frac{\left(2-3 b \right)^{2}}{48\pi \left(9 b^{2}-24 b +8\right) \left(1-b \right) \,Q^{2}}, \\ 
T_{c} &=& \frac{2 \left(2-3 b\right)^{2} \sqrt{6}}{9\pi \sqrt{\left(2-3 b \right) \left(9 b^{2}-24 b +8\right)} \, \left(4-3 b \right) Q }, \\
v_{c} &=& \frac{\sqrt{6}\, \sqrt{\left(2-3 b \right) \left(9 b^{2}-24 b +8\right)}\, Q}{2-3 b }.
\end{eqnarray}
The critical triple $(P_{c}, T_{c}, v_{c})$ provides the following ratio
\begin{equation}
\chi = \frac{P_c v_c}{T_c} = \frac{3}{8} + \frac{3 b}{32} + O\left(b^{2}\right),
\end{equation}
where certain approximations have been used.  This ratio  is greater  than that of the Van der Waals fluid 
systems since $b$ is a positive quantity.  Taking $b = 0$, the critical quantities can be reduced to 
\begin{eqnarray}
P_c &=& \frac{1}{96 \pi Q^2}, \\ 
T_c &=& \frac{\sqrt{6}}{18 \pi Q}, \\
v_c &=& 2 \sqrt{6}\, Q,
\end{eqnarray}
recovering the critical thermodynamic variables for charged RN-AdS black holes \cite{7}. In this case, the critical triple $(P_c, T_c, v_c)$ provides the following ratio
\begin{equation}
\chi = \frac{P_c v_c}{T_c} = \frac{3}{8}.
\end{equation}
This shows  that   the Van der Waals compressibility ratio $\chi$ can be recovered by sending $b$ to zero \cite{39}. To support such   critical properties,  we illustrate the   $P$-$V$
diagram, as shown in Fig. (\ref{F41}).

At temperature $T$ above the critical value $T_c$, the system behaves like an ideal gas. in this way, the critical isotherm at $T = T_c$ is indicated by an inflection point at the critical volume $v_c$ and the critical pressure $P_c$. At $T < T_c$, there is an unstable thermodynamic region. The $P$-$V$ diagram clearly resembles that of a Van der Waals fluid. In addition, the new  NC parameter $b$ influences the thermodynamic behaviors of the  studied system. Indeed, when $b$ increases, the minimum value of the  pressure $P$ also decreases for the same temperature $T$. This reveals that  such a  NC parameter  has the effect of modifying the structure of the $P$-$V$ diagram.

\subsection{Phase transitions}
\begin{figure}[h!]
    \centering
    \begin{tabular}{cc}
       \includegraphics[width=7cm,height=7cm]{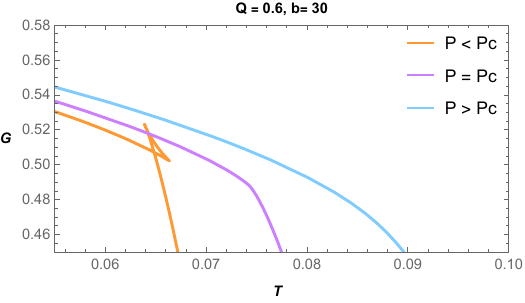} & 
         \includegraphics[width=7cm,height=7cm]{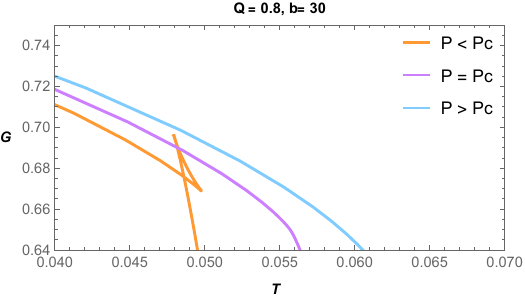} 
    \end{tabular}
    \begin{tabular}{cc}
       \includegraphics[width=7cm,height=7cm]{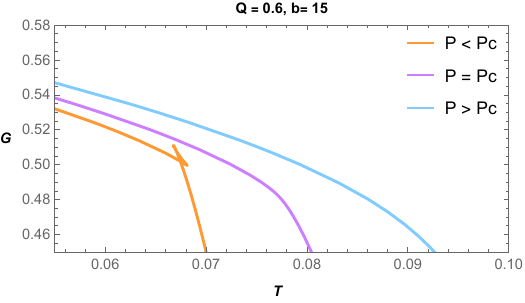} & 
         \includegraphics[width=7cm,height=7cm]{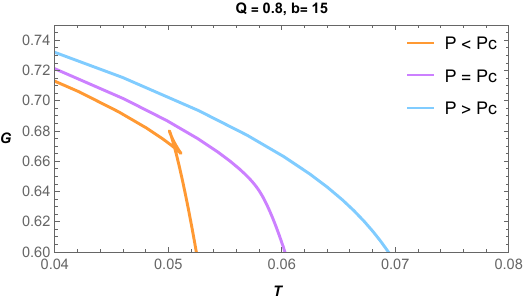} 
    \end{tabular}
    \caption{Gibbs free energy in terms of the temperature for different values of $P$ and $Q$.}
    \label{7}
\end{figure}

Here, we discuss  the phase transitions by approaching   the Gibbs free energy  given by  
\begin{equation}
   G = M - T_H S.
\end{equation}
The computations give
{\footnotesize
\begin{equation}
G = \frac{16 P \pi  \,r_h^{5}(r_h-2a)-3 Q^{2}(9r_h^2-8ar_h+a^2)-6 r_h^{3}(r_h+a)}{12r_h^{2} \left(a-2 r_h  \right) }.
\end{equation}
}
Taking $Q=0$, the Gibbs free energy reduces to
\begin{equation}
G = \frac{r_{h} \left( 3r_{h} - 8\pi P r_{h}^{3} + a \left( 3 + 16\pi P r_{h}^{2} \right) \right)}{6 \left( 2r_{h} - a \right)},
\end{equation}
recovering  the result obtained  in \cite{30}.
 Exploiting  the critical   thermodynamic quantities, the $G-T$ diagrams  are  presented  in Fig.(\ref{7}).

The $G-T$ curves, showing the Gibbs free energy as a function of temperature, have similar shapes for different values of the critical pressure $P_c$. 
For pressures below the critical value ($P < P_c$), the diagram shows a swallow-tail shape. 
This shape signals a first-order phase transition between small and large black holes. 
Increasing the electric charge $Q$ or changing the NC  parameter $b$ moves the curves and changes their shape, affecting the temperature and Gibbs free energy at the phase transitions. 
This behavior is very similar to Van der Waals fluids, supporting the analogy between  such NC black holes and classical fluids.

\section{Joule-Thompson expansion}
To unveil extra  thermodynamic data, we  examine    the Joule-Thompson expansion  developed in \cite{40,41,42}.   It is recalled that  the Joule-Thomson coefficient  reads as 
\begin{equation}
\mu=\left( \dfrac{\partial T}{\partial P} \right)_{M}=\dfrac{1}{C_{P}} \left[ T \left( \dfrac{\partial V}{\partial T} \right)_{P}-V  \right]. \label{m}
\end{equation}
To approach such an expression,   the equation of state  in  terms of the thermodynamic volume will be needed.
Considering Eq.(\ref{V}), Eq.(\ref{P}) and Eq.(\ref{xxx}), we  can find  the temperature as a function  of the volume and the pressure. Indeed,  we have 
\begin{equation}
T = \dfrac{1}{2 V  (3b-4)} \left(  
16 P V (b-1)\left( \dfrac{3V}{4\pi}\right) ^{\!1/3} +4\left( b-\dfrac{2}{3}\right) \left( \dfrac{3V}{4\pi}\right) ^{\!2/3} +3Q^2 \left(b^2-\dfrac{8}{3}b+\dfrac{8}{9} \right)  \right). \label{t}
\end{equation}
 Using Eq.(\ref{t}) and the second part of Eq.(\ref{m}), we can obtain the temperature associated with a zero Joule-Thomson coefficient.  The computations reveal that  the   repeated inversion temperature $T_i$ is given by 
\begin{equation}
T_i = \dfrac{1}{6 V  (3b-4)} \left(  
16 P V (b-1)\left( \dfrac{3V}{4\pi}\right) ^{\!1/3} -4\left( b-\dfrac{2}{3}\right) \left( \dfrac{3V}{4\pi}\right) ^{\!2/3} -9Q^2 \left(b^2-\dfrac{8}{3}b+\dfrac{8}{9} \right)  \right).  \label{tt}
\end{equation}
Exploiting  the volume quantity,  this temperature   can be expressed as \begin{equation}
T_i= \dfrac{64P \pi  (b-1) r_h^4 -12(b-\dfrac{2}{3})r_h^2-27Q^2 \left(b^2-\dfrac{8}{3}b+\dfrac{8}{9} \right)}{24\pi (3b-4) r_h^3} .
\label{ti1}
\end{equation}
By help of  Eq. (\ref{t}),  we find 
\begin{equation}
T= \dfrac{64P \pi  (b-1) r_h^4 +12(b-\dfrac{2}{3})r_h^2+9Q^2 \left(b^2-\dfrac{8}{3}b+\dfrac{8}{9} \right)}{8\pi (3b-4) r_h^3} .\label{ti2}
\end{equation}
Subtracting Eq. (\ref{ti1}) form Eq. (\ref{ti2}),  we  obtain an algebraic equation given by
 \begin{equation}
64 P_i\pi \mathit{Ar_i}^{4}+24 Br_i^2 +27 Q^{2} C  = 0, 
 \end{equation}
where one has used 
\begin{equation}
\begin{aligned}
A=&b-1,\\
B=&b-\dfrac{2}{3}, \\
C=& b^2-\dfrac{8}{3}b+\dfrac{8}{9}.
\end{aligned}
\end{equation}
In this equation,  $P_{i}$  indicates the inversion pressure.  Leaving only  the real and positive root, we  obtain 
 \begin{equation}
r_i = \frac{\sqrt{3}\,  \left( B-\sqrt{B^{2}-12 P_i\pi  A \,Q^{2} C }\right)}{4  \sqrt{P_i\pi  A}}.
\end{equation}
By inserting this root into Eq.(\ref{ti1}), the inversion temperature  is found to be 
\begin{equation}
T_{i}=\frac{2 \sqrt{3}\, \left(A \! K^{2}+4 \pi  \,Q^{2} A^{2} C +\frac{\mathit{ABK}}{P_i}\right)}{\sqrt{P_i\pi  A K}\, \pi  \left(3 b-4 \right)K},
\end{equation}
where one used $K=\sqrt{B^{2}-12 P_i\pi  A \,Q^{2} C }-B$. At zero inversion pressure $P_i=0$, the inversion temperature comes to  its minimum value
 \begin{equation}
T_{i}^{min}=\frac{\sqrt{6}\left(3 b -2\right)^{2} }{9\pi\sqrt{ \left(2-3 b \right) \left(9  b^{2}-24 b +8\right)}\, \left(3 b -4\right) Q}. 
 \end{equation}
This produces a relationship between the minimum inversion and critical temperatures being the following ratio 
 \begin{equation}
\xi= \dfrac{T_{i}^{min}}{T_{c}}=\dfrac{1}{2}.
 \end{equation}
This reveals that  the  obtained  result   matches  perfectly with the charged AdS black hole universal behaviors with respect to the electric  charge $Q$ and the NC parameter $b$ \cite{43,430}.  This supports the validity  of the proposed black hole metric.

\begin{figure}[h!]
\centering
\begin{tabular}{cc}
\includegraphics[width=6.5cm,height=6.5cm]{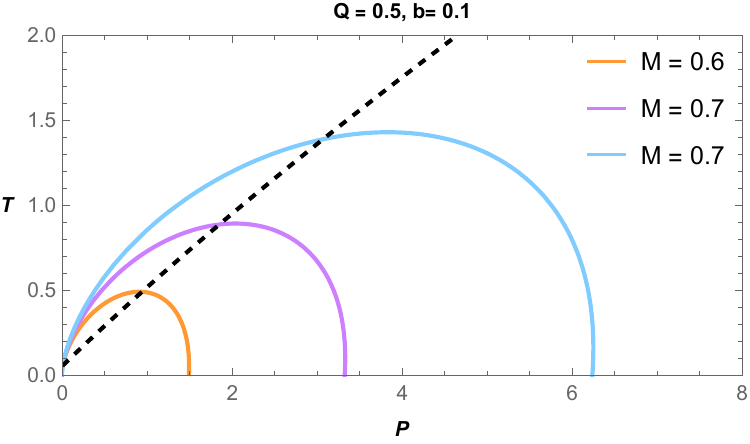} &
\includegraphics[width=6.5cm,height=6.5cm]{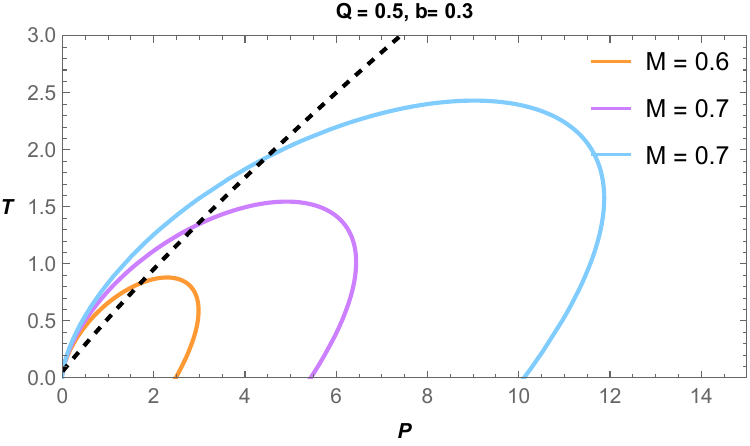}
\end{tabular}
\begin{tabular}{cc}
\includegraphics[width=6.5cm,height=6.5cm]{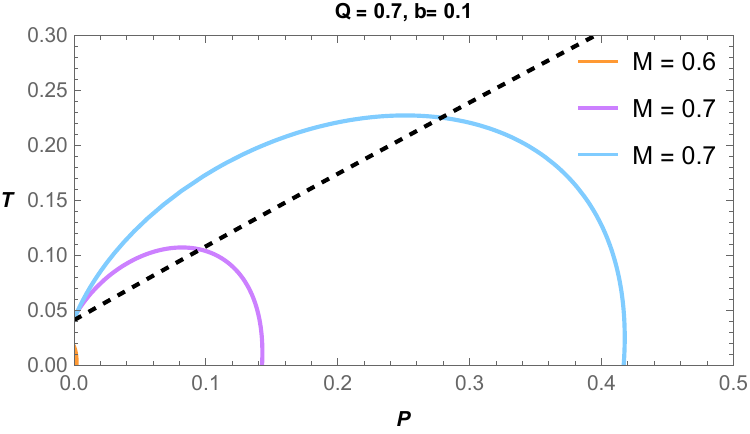} &
\includegraphics[width=6.5cm,height=6.5cm]{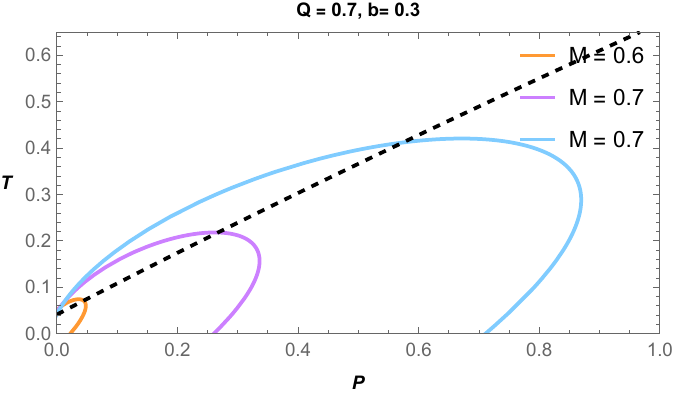}
\end{tabular}
\caption{Inversion (dashed lines) and isenthalpic (solid lines) curves for Noncommutative RN-AdS black holes, for different values of $M$, $b$ and $Q$.}
\label{55}
\end{figure}
Fig.(\ref{55}) reveals that the inversion curves separate the $(T, P)$ diagram into two distinct regions.  Over the inversion curves, the system cools, while under them, it is warming. 
This can be seen from the slope of the isenthalpic curves. Indeed,  a positive slope means cooling,  and a negative slope means warming behaviors. At the inversion curve itself, there is neither warming nor cooling marking  the boundary between the two regimes.

\newpage
 
\section{Conclusions}
In this work, we have explored the Reissner–Nordström–AdS black holes in noncommutative spacetime with Lorentzian-smeared distributions. Subsequently, we have investigated the thermodynamical properties of charged AdS black holes within the framework of noncommutative spacetime. Specifically, we have analyzed the thermal stability and  the critical behaviors, including phase transitions. In particular, we have calculated the relevant thermodynamic quantities needed to approach certain physical  behaviors. By applying the associated laws, we have evaluated the heat capacity to assess the stability of the black holes and have identified the regions where they remain stable. By relating the NC parameter $a$ to the horizon radius $r_h$ through a constant parameter $b$, we have studied the $P$--$V$ criticality. In particular, we have determined the critical pressure $P_c$, the critical temperature $T_c$, and the  critical specific volume $v_c$ in terms of $b$. We have shown that the ratio $\dfrac{P_c v_c}{T_c}$ represents a universal number independent of the charge $Q$. In the small-limit regime of the external parameters, we have recovered behaviors analogous to those of Van der Waals fluids. Then, we have examined the phase transitions by computing  and analyzing the Gibbs free energy variations. Finally, we have investigated the Joule--Thomson expansion for these black holes and have revealed the similarities and differences with Van der Waals fluids. This universal behavior supports the validity of the proposed  black  hole metric  in noncommutative spacetime. \\
This work has raised several open questions. A natural extension would be to explore other properties, including optical features, such as the shadow and  the light deflection near these NC  black holes,  where a possible contact with  M87$^*$ and SgrA$^*$ bands could be elaborated \cite{WM}.

  {\bf Data availability}\\
  Data sharing is not applicable to this article.

\section*{Author contributions}
 Manuscript writing: All authors\\
  Data analysis and interpretation: All authors\\
  Final approval of manuscript: All authors

\section*{Acknowledgements}
The authors  would like to thank  A. Belhaj for  useful suggestions,  discussions, communications,   and encouragement. They also thank the anonymous referees for their helpful comments, suggestions, and
 remarks that improved the quality of the manuscript.   MJ would like to thank S.E. Baddis and H. Belmahi for collaborations on related topics.
This work was done with the support of the CNRST in the frame of the PhD Associate Scholarship Program PASS.

\end{document}